\documentclass[aps,prb, twocolumn,groupedaddress,showpacs]{revtex4}
\usepackage{graphicx}

\begin{document}

\title{Exchange energy and generalized polarization in the presence of 
spin-orbit coupling in two dimensions}
\author{Stefano Chesi}
\author{Gabriele F. Giuliani}
\affiliation{Department of Physics, Purdue University,
West Lafayette, IN 47907, USA}

\date{\today}

\begin{abstract}
We discuss a general form of the exchange energy for a homogeneous system of interacting 
electrons in two spatial dimensions which is particularly suited in the presence of 
a generic spin-orbit interaction. The theory is best formulated in terms of a generalized
fractional electronic polarization. Remarkably we find that a net generalized polarization 
does not necessarily translate into an increase in the magnitude of the exchange energy, 
a fact that in turn favors unpolarized states. Our results account qualitatively for the 
findings of recent experimental investigations.
\end{abstract}

\pacs{71.10.Ca, 71.10.-w, 71.70.Ej, 73.61.Ey}

\maketitle

In most cases, as a first step in assessing the role of the electronic interactions,
an analysis of the role played by the exchange energy is illuminating and most 
often very useful. As we show here, the exchange energy acquires particularly interesting 
and unusual properties in two dimensions in the presence of spin-orbit of the 
Rashba type. In this case the chiral nature of the single particle states involved competes 
with standard requirements that are commonly identified with the exchange interaction.

The interplay of spin-orbit coupling and Coulomb interaction is unexplored, except for 
the initial and limited work of Refs.~\onlinecite{chen99,saraga05} where some quasi-particle 
properties were investigated. Most of the work is relevant in the high density limit in 
which only weak corrections to the standard behavior were identified. Systematic studies of 
the mean-field phase diagram of this interesting system are provided 
elsewhere.\cite{gfg_proc05,unpub_ChesiGFG,unpub_SimionGFG_inhom}

Alongside the general theoretical relevance of such systematic studies, a timely practical 
application of such an analysis can be found in the study of spin polarization states 
in two dimensional hole systems in a $GaAs$ heterostructure.\cite{winkler05} 

The expression for the exchange energy per particle in a homogeneous electron liquid is 
well known.\cite{TheBook} In particular, for the two dimensional case, which is of 
relevance in the present discussion, the formula is given by:
\begin{equation}\label{exch0}
\mathcal{E}_{x,0}(r_s, p) = - 
\frac{4\sqrt{2}}{3\pi}\frac{(1+p)^\frac{3}{2}+(1-p)^\frac{3}{2}}{r_s} ~,
\end{equation}
where we measure energy in Rydbergs and the significance of the sub index $0$ will be 
presently made clear. 

In Eq.~(\ref{exch0}) the parameter $p$ is defined as follows:
\begin{equation}\label{p}
p=\frac{n_+-n_-}{n}~,
\end{equation} 
where $n$ is the particle number density and $n_\pm$ refers to the particle density corresponding 
to the two spin subbands. $r_s=1/\sqrt{\pi a_B^{2} n}$ is the usual density parameter in two dimensions. 
It is important to notice that Eq.~(\ref{exch0}) is obtained under the assumption that the spins 
are quantized along a common arbitrary quantization axis $\hat{z}$. 
Therefore, when, as in this particular case, the label $\pm$ refers to the orientation 
$\uparrow\downarrow$ of the spin, $p$ assumes the meaning of a fractional spin polarization.
It is easy to see that the exchange energy (\ref{exch0}) monotonically attains 
its maximum magnitude for $p=1$, and therefore favors the polarization of the system. In fact, within the 
Hartree-Fock approximation, a transition to a fully spin polarized 
state occurs for $r_s=\frac{3\pi}{8(2-\sqrt{2})}\simeq 2.01$.\cite{TheBook}

As we will presently show, this scenario changes in interesting ways in the presence of the 
spin-orbit interaction. In the quasi two dimensional electronic systems present in a high 
quality semiconducting heterostructure, the latter can come in different guises ranging from 
Rashba and Dresselhaus couplings to terms induced by the application of an external 
magnetic field.\cite{WinklerSpringer03} 
The following is a systematic study of the effects of spin-orbit interactions of various 
type on the exchange energy and the polarization properties of a clean two dimensional system. 
The theory can be elegantly formulated in a compact way by introducing a generic form of 
spin orbit coupling, something that allows us to draw quite general conclusions.

The paper is organized as follows: Section \ref{modH} introduces our model hamiltonian
including a spin-orbit coupling of generic form; Section \ref{nonint} describes the solution 
of the corresponding non interacting problem in terms of the generalized polarization. 
In Section \ref{int}, which represents the main part of the paper, the role of the 
exchange energy is studied in detail. In particular, it is demonstrated there that the 
presence of spin-orbit coupling leads to a qualitatively different relation
between this quantity and the generalized polarization.
Section \ref{exp} contains a discussion of recent experiments on the polarization of
holes in $GaAs$ heterostructures in the light of these findings.
Finally some general remarks and conclusions are provided in Section \ref{disc}. 

\section{Model hamiltonian}
\label{modH}
We consider the following generic model two dimensional electronic system 
described by the hamiltonian\cite{comm_neutralizingbackground}
\begin{equation}\label{Htot}
\hat H_n ~=~ \sum_i \hat H^{(i)}_{0,n}+\frac{1}{2}\sum_{i\neq j}
\frac{e^2}{| \hat{\bf r}_i- \hat{\bf r}_j|} ~,
\end{equation}
where the single-particle terms are of the form:  
\begin{equation}\label{H0}
\hat H_{0,n}=\frac{\hat \mathbf{p}^2}{2 m}+i \gamma \, 
\frac{(\hat p_-)^n \hat\sigma_+ - (\hat p_+)^n \hat\sigma_- }{2} ~,
\end{equation}
the motion taking place in the x-y plane. In Eq.~(\ref{H0}) we have defined 
$\hat{p}_\pm=\hat p_x\pm i \hat p_y $ and 
$\hat{\sigma}_\pm=\hat \sigma_x\pm i \hat \sigma_y $. 
Here $n$ is an integer number assuming values from $0$ to $3$. 
In view of its structure we will refer to the second term in (\ref{H0}) as a 
generalized spin-orbit coupling.

In the simplest case, $n=0$ corresponds to the familiar Zeeman coupling 
with $\gamma = \frac{g \mu_B B}{2}$ for which the standard result (\ref{exch0}) applies. 
More complicated $n=0$ expressions are contemplated in the Appendix.

For $n>0$, Eq.~(\ref{H0}) describes different types of bona fide spin-orbit interactions. 
In particular, for $n=1$ we obtain a form equivalent to the Rashba\cite{bychkov84a,bychkov84b} or 
Dresselhaus\cite{dresselhaus55} spin-orbit hamiltonians, appropriate for two-dimensional 
conduction electrons in III-V semiconductors.
Furthermore, the cases $n=2$ and $n=3$ are appropriate for holes in III-V semiconductors 
as $GaAs$. Explicit expressions are provided in the Appendix. 

From a physical point of view the $n=2$ term arises in the presence of an external in-plane 
magnetic field, the coefficient $\gamma$ being proportional to the value $B$ of such field. 
For hole carriers in $GaAs$ heterostructures, this term is the dominating low-field effect 
in the high symmetry growth directions $[001]$ and $[111]$, for which the Zeeman term can be 
safely assumed to be approximately vanishing.\cite{WinklerSpringer03}

Finally, the $n=3$ term corresponds to a Rashba spin-orbit coupling for holes, the 
value of $\gamma$ being in this case approximately proportional to the value of 
the average electric field of the confining potential.\cite{commExchange1}

\section{Non interacting electrons}
\label{nonint}
The eigenstates of (\ref{H0}) are plane waves, a fact that allows one to write the spin-orbit 
term in the form $\gamma \, (\hbar k)^n \, \mbox{\boldmath $\sigma$}\cdot \hat{s}_\mathbf{k}$, 
where $\hat s_{\bf k}$ is defined as follows:
\begin{equation} \label{sk}
\hat{s}_{\bf k}= -\sin(n\phi_{\bf k}) \hat x + \cos(n\phi_{\bf k}) \hat y  ~.
\end{equation}
Here $\phi_{\bf k}=\arctan(k_y/k_x)$ is the polar angle spanned by $\mathbf{k}$. 
The unit vector $\hat{s}_{\bf k}$ determines the direction of the quantization axis for the 
particular value of the wave vector $\mathbf{k}$. The two possible spin orientations immediately 
give the eigenstates:
\begin{equation}
\label{phi0kpm}
\varphi_{ {\bf k} , \pm} ( {\bf r} ) ~=~
\frac{e^{i {\bf k} \cdot {\bf r} }}{\sqrt{2 L^2}}
\left(
\begin{array}{c}
\pm 1\\
i e^{i n \phi_{\bf k}}
\end{array}
\right) ~,
\end{equation}
with energies:\cite{commExchange2}
\begin{equation}\label{energies}
\epsilon_{\mathbf{k},\pm}=\frac{\hbar^2 k^2}{2 m}\mp \gamma (\hbar k)^n  ~.
\end{equation} 

Notice that in view of the structure of the spinors the integer $n$ can be seen as a 
spin direction winding number.

It is not difficult to show\cite{gfg_proc05} that a class of determinantal 
many-body states with compact momentum space occupation which are 
homogeneous and isotropic in the plane of motion are uniquely 
determined by the areal density $n$ and the fractional polarization $p$ 
defined as in (\ref{p}).\cite{commExchange4,unpub_SimionGFG_inhom}
It must be immediately noted here that in the general case $p$ should not be interpreted as 
a spin polarization for this is in general vanishing in these states. $p$ does merely 
determines the two Fermi vectors $k_\pm$ of the {\it so called} spin-split subbands via the 
relations:
\begin{equation}\label{kpm}
k_\pm=\sqrt{2\pi n(1\pm p)} ~,
\end{equation}
which uniquely determine the occupation numbers $n_{\bf k \pm}$ of the subbands.

It is readily found that the non interacting energy per particle (in Rydbergs) of 
such Slater determinants is given by:
\begin{equation}\label{en0}
\mathcal{E}^{(0)}_{n}(r_s,p)=\frac{1+p^2}{r_s^2}-\bar\gamma
 \,\frac{2^{\frac{n}{2}}}{r_s^n} \, 
 \frac{(1+p)^{1+\frac{n}{2}} -(1-p)^{1+\frac{n}{2}}}{1+\frac{n}{2}}  ~,
\end{equation}
where we have defined the dimensionless coupling:
\begin{equation}\label{gamma}
\bar\gamma=\frac{m^{n-1}e^{2(n-2)}}{\hbar^{n-2}}\,\gamma  ~.
\end{equation}
The first term in (\ref{en0}) is the contribution of the kinetic energy while the 
second corresponds to the spin-orbit interaction.

In this case, minimization of the total energy with respect to the value of $p$,
$\frac{\partial \mathcal{E}^{(0)}_{tot}}{\partial p}=0$, leads to the (finite)
ground state value of the equilibrium generalized polarization $p_{min}^{(0)}$.
Interestingly the result only depends on the dimensionless quantity 
\begin{equation}\label{gdef}
g ~= ~ 2^{\frac{n}{2}}\bar\gamma r_s^{2-n} ~, 
\end{equation}
so that it can be compactly expressed as follows:
\begin{equation}\label{pmin}
p_{min}^{(0)}=\left\{
\begin{array}{cl}
g  &  \mathrm{for} \,\, n=0 ~,  \\
g \sqrt{1-\frac{g^2}{4}}  &  \mathrm{for} \,\, n=1 ~,\\
g     &  \mathrm{for} \,\, n=2  ~, \\
g \sqrt{
\frac{-3 g^4+6 g^2 - 2+2
(1-2 g^2)^{3/2}}{g^6}}  & \mathrm{for}\,\,  n=3 ~.
\end{array}
\right.
\end{equation}
For small coupling this quantity behaves as 
\begin{equation}\label{pmin0smallg}
p_{min}^{(0)} ~\simeq~ g +\frac{n(n-2)}{8} g^3 + O(g^5)~. 
\end{equation}
One can then immediately notice that since for $n=1$ we have $g=\sqrt{2}\bar\gamma r_s$, 
in this case the high-density regime is equivalent to a vanishing spin-orbit. 
The opposite obtains for $n=3$, since $g=2\sqrt{2}\bar\gamma/r_s$. For quadratic 
spin-orbit $g=2\bar\gamma$, and the fractional generalized polarization $p$ is 
independent of the density.

It is also useful to define here the depopulation coupling strength 
$\bar \gamma^{(0)}_d$ as the particular value of $\bar \gamma$ for which, 
at a given density, the upper band empties and $p_{min}^{(0)}=1$. This is readily 
obtained from (\ref{pmin}) to be given by:
\begin{equation}\label{gd_nonint}
\bar\gamma_d^{(0)}=
\left\{
\begin{array}{cl}
\frac{1}{r_s^2}  &\qquad {\rm for}~n = 0 ~,\\
\frac{r_s^{n-2}}{2^{n-1}}  &\qquad {\rm for}~n =1,2,3 ~~.
\end{array}
\right. 
\end{equation}

\section{Effects of exchange}
\label{int}
We turn next to the effect of exchange. If translational invariance is not 
broken,\cite{unpub_SimionGFG_inhom} a single Slater determinant can be constructed 
with plane waves states characterized by generic orientation of the spin quantization 
axis $\hat s_{\bf k}$ and occupation numbers $n_{\mathbf{k}\pm}$. The total exchange 
energy can be written in the following elegant general form:\cite{unpub_ChesiGFG,gfg_proc05}
\begin{equation}\label{Ex}
E_{x}
= -\frac{1}{2 L^2}\sum_{\mathbf{k},\mathbf{k}';\mu ,\mu' = \pm} v_{\mathbf{k}-\mathbf{k}'}\,
\frac{1+ \mu \mu'\,\, \hat{s}_{\mathbf{k}}\cdot \hat{s}_{\mathbf{k}'}}{2}\,
n_{\mathbf{k}\,\mu}n_{\mathbf{k}'\mu'} ~, 
\end{equation}
which represents a functional of $n_{\mathbf{k}\,\pm}$ and $\hat{s}_{\mathbf{k}}$.
Eq.~(\ref{Ex}) immediately reduces to the familiar textbook result when $\hat s_{\bf k}=\hat z$. 
In this case only states with parallel spin contribute.

Making use of Eq.~(\ref{sk}) in (\ref{Ex}) leads to the following result for the 
exchange energy per particle (in Rydbergs):

\begin{equation}\label{SOexch}
\mathcal{E}_{x,n}(r_s,p)=\mathcal{E}_{x,0}(r_s,p)+\delta\mathcal{E}_{x,n}(r_s,p) ~,
\end{equation}
where the correction to Eq.~(\ref{exch0}) can be obtained from the following quadrature:
\begin{widetext}
\begin{eqnarray}\label{Dexch}
\delta\mathcal{E}_{x,n}(r_s,p)=\frac{\sqrt{2}}{4\pi r_s}
\int_{\sqrt{1-p}}^{\sqrt{1+p}} x \,{\rm d}x 
\int_{\sqrt{1-p}}^{\sqrt{1+p}} y \,{\rm d}y  
\int_{0}^{2\pi}\frac{1-\cos n\theta}{\sqrt{x^2+y^2-2xy\cos{\theta}}} \,
{\rm d}\theta  ~.
\end{eqnarray}
\end{widetext}
 
\begin{figure}
\begin{center}
\includegraphics[width=0.35\textwidth]{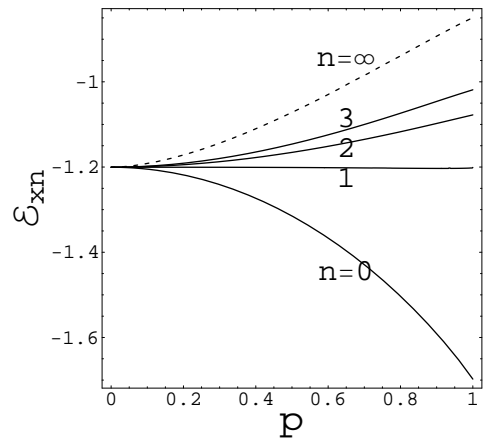}
\caption{\label{exchplot} Plot of the exchange energy per particle $\mathcal{E}_{xn}(r_s,p)$ 
(in $Ry$ units) as a function of $p$. Here $r_s=1$. The different values of $n$ are noted,
and the limiting curve $n=\infty$ is also displayed (dashed line).}
\end{center}
\end{figure}

The resulting exchange energy is plotted in Figure \ref{exchplot} as a function of $p$. 
We notice that for $p=0$, the exchange energy is independent of the spin 
quantization axis orientations $\hat s_\mathbf{k}$. This can be understood by realizing 
that the corresponding many-body state can be constructed by repeated application of 
$b_{\mathbf{k}+}^\dag b_{\mathbf{k}-}^\dag$, an operator that creates a spin singlet and 
is therefore independent of the spin quantization direction. As a consequence the $\hat s_{\bf k}$ 
dependence of all the physical quantities (e.g. the exchange and the spin orbit energies) 
stems only from the existence of regions of momentum space where $n_{\mathbf{k}+}\neq n_{\mathbf{k}-}$.
It is also important to remark that only for $n=0$ is the magnitude of the exchange energy maximum for
$p=1$. For the other cases the minimum occurs at $p = p_1^* \simeq 0.915$ (although this is not obvious 
from Figure \ref{exchplot}) for $n=1$ and at $p=0$ for $n \geq 2$.

The different behavior of the exchange energy in the various cases leads to dissimilar 
results.

For $n=0$, the fact that the minimum of $\mathcal{E}_{x,0}$ occurs at $p=1$ leads to 
an enhancement of $p_{min}$ i.e. to the familiar enhancement of the spin polarization. 
The opposite is true for $n \geq 2$ since in these cases the minimum of $\mathcal{E}_{x,0}$ 
occurs at $p=0$. 

For completeness we remark that in the limiting case of very large winding number $n$ 
the result can be obtained simply neglecting the $\cos n\theta$ contribution 
in Eq.~({\ref{Dexch}). For comparison the corresponding curve is shown as a dashed 
line in Figure \ref{exchplot}. Notice that in this case the magnitude of the exchange 
energy is minimum for $p=1$, the value being given by:
\begin{equation}
\lim_{n\to \infty}\mathcal{E}_{x,n}(r_s,1)=-\frac{8}{3\pi r_s} ~.
\end{equation}

The situation for the $n=1$ case is more complex although in all cases the exchange 
only leads to a very small deviation from $p_{min}^{(0)}$. Specifically $p_{min}$ is 
slightly enhanced for $p^{(0)}_{min} < p_1^*$ and slightly diminished for 
$p^{(0)}_{min} > p_1^*$, being unrenormalized for $p^{(0)}_{min} = p_1^*$. 

A similar argument leads one to conclude that the critical value $\bar\gamma_d$ for 
which the upper spin band empties (at fixed $r_s$) decreases from its non interacting 
value for $n=0$ ($g=1$) while it does increase in the other cases. 

Studying the limit of small $p$ is of particular interest since it corresponds to 
a determination of the generalized susceptibility. In this case a direct inspection of 
the integral of Eq.~(\ref{Dexch}) leads to the asymptotic formula:
\begin{equation}\label{Ex_smallp}
\mathcal{E}_{x,n}(r_s,p)\simeq -\frac{8\sqrt{2}}{3\pi r_s} 
- \frac{C_n}{r_s} \, p^2  ~,
\end{equation}
where we have defined the quantity
\begin{equation}\label{Cn}
C_n ~ = ~ \frac{\sqrt{2}}{\pi}\sum_{m=0}^n \frac{1}{1-2m} ~.
\end{equation}

The resulting value for $p_{min}$ is then given by:
\begin{equation}\label{small_pmin}
p_{min}\simeq \frac{g}{1-C_n r_s}~.
\end{equation}

Eq.~(\ref{small_pmin}) simply expresses the fact that in this limit the effect of the interactions
is to renormalize the non interacting result $p_{min}^{(0)} \simeq g$ via the denominator 
$(1-C_n r_s)^{-1}$. Interestingly the latter corresponds to an enhancement only for $n=0$.

In particular for $n=0$ we recover the well known Hartree-Fock differential instability occurring
at $r_s=\frac{\pi}{\sqrt{2}}$.\cite{TheBook} On the other hand, for $n=1$ we have $C_1=0$ the whole
renormalization effect being solely associated with correlations effect. 
Finally for $n \geq 2$, $C_n$ is (ever increasingly) negative leading to a perhaps iconoclastic 
exchange driven quenching of the generalized polarization. 

\begin{figure}
\begin{center}
\includegraphics[width=0.4\textwidth]{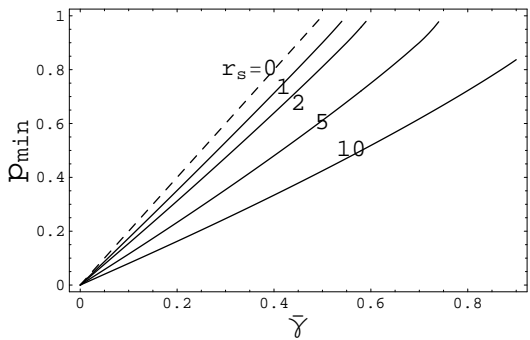}
\caption{\label{pol} Plot of the fractional generalized polarization $p_{min}$ as 
function of the parameter $\bar\gamma$ for different values of $r_s$. Here $n=2$. 
The increase of the depopulation value
$\bar\gamma_d$ with $r_s$ is manifest.}
\end{center}
\end{figure}

Expression (\ref{small_pmin}) is valid in the limit of $\bar\gamma \to 0$ or, for $n>1$, 
when $r_s$ is large. The generic case when $p_{min}$ is not small must be 
obtained numerically. As an example we show in Figure \ref{pol} the value of $p_{min}$ 
as function of the dimensionless coupling strength $\bar\gamma$ for different values of the 
density parameter $r_s$ in the case of $n=2$. We notice that for low densities the 
depopulation value $\bar\gamma_d$, for which  $p_{min}=1$, is considerably 
increased by exchange. This is in stark contrast with the familiar $n=0$ case. 
In particular, for the $n=2$ case of Figure \ref{pol} we have: 
\begin{equation}\label{gd_inter}
\bar\gamma_d=\frac{1}{2}+
\left(\frac{1}{12}-\frac{1}{9\pi} \right) \, r_s ~,
\end{equation}
where the second term represents the increase of the depopulation field due to exchange
effects. In this particular case the correction is linear in the density parameter $r_s$.

\section{Application to spin polarized hole systems}
\label{exp}
Spin polarization experiments have recently been performed on $GaAs$ two dimensional 
hole systems with growth direction along $[113]$ and $[100]$.\cite{winkler05}  
In these studies, the magnitude of the in-plane depopulation field $B_d$, when only 
one band is occupied, is surmised from the measured longitudinal magneto-resistance. 
While a small suppression of $B_d$ with respect to its non interacting value $B_d^0$ 
is observed in the $[113]$ case, basically no suppression is observed for the $[100]$ 
growth direction. This must be contrasted with the fact that, as discussed in 
Ref.~\onlinecite{winkler05}, in the absence of spin-orbit interaction and at the 
densities under consideration, the expected ratio for electrons is of 
order $B_d^0/B_d \sim 10$.

The experiments of Ref.~\onlinecite{winkler05} are carried out at somewhat low densities 
($r_s \simeq 10-15$) a regime in which a simple Hartree-Fock treatment, as 
well as somewhat more sophisticated analytical treatments designed to approximately 
include correlation effects, generally fail to provide reliable quantitative results. 
On the other hand, Monte Carlo analysis of the system being to date non existent, it is 
reasonable to expect that many of the qualitative features established 
by a study of the exchange energy will prove sufficiently robust to justify such 
a preliminary discussion.

In these quantum wells the introduction of a magnetic field induces in the effective 
hamiltonian of two-dimensional holes the spin dependent terms given in Eqs.~(\ref{so2}), 
(\ref{zeeman3}), and (\ref{zeeman1}) of the Appendix. All these terms can be reduced 
to some of the generic forms contemplated by our model hamiltonian (\ref{H0}).
The corresponding coupling strength $\gamma$ for each of these terms can then be
extracted, as for instance  explicitly done in (\ref{gamma21}), (\ref{gamma03}), and
(\ref{gamma01}).
One can then make use of the definitions (\ref{gamma}) and (\ref{gdef}), alongside 
suitable numerical parameters, to determine the relevant dimensionless coupling strength. 

For our numerical estimates we use values appropriate for the case of 
Ref.~\onlinecite{winkler05}. In particular: $W=200$~\AA~for the width of the 
quantum well, $m\simeq 0.2 \, m_0$ for the effective mass, $\epsilon=12.4$ 
for the background dielectric constant, and $n=3~10^{10}~{\rm cm}^{-2}$ for the hole 
density (corresponding to $r_s \simeq 10$). 
 
We first consider the $[113]$ growth direction. In this case the various physical properties
are anisotropic in the plane, the principal directions being given by $x \equiv [1 \bar 1 0]$ 
and $y \equiv [33\bar 2]$. For the density under consideration the depopulation field is 
approximately given by $B_d \simeq 10 ~ {\rm T}$ along $x$ and $B_d \simeq 5 ~ {\rm T}$ along $y$. 
This gives the following results for various dimensionless couplings: 
For Eq.~(\ref{so2}) we have
\begin{equation}\label{strength21}
|g^x_{21}|= 0.19  \quad {\rm and} \quad |g^y_{21}|= 0.09~;
\end{equation}
For Eq.~(\ref{zeeman3}):
\begin{equation}\label{strength03}
|g^x_{03}|=0.23  \quad {\rm and} \quad |g^y_{03}|=0.03  ~;
\end{equation}
Finally for Eq.~(\ref{zeeman1}):
\begin{equation}\label{strength01}
|g^x_{01}|=0.14  \quad {\rm and} \quad |g^y_{01}|=0.16  ~.
\end{equation}
We recall here that (\ref{so2}) is a term of type $n=2$, while (\ref{zeeman3}) 
and (\ref{zeeman1}) are both of type $n=0$. We notice that the quadratic spin-orbit,
although in general smaller, has a strength which is comparable to that of the terms 
of the Zeeman type. 

For the $[100]$ growth direction, the Zeeman term (\ref{zeeman1}) is vanishing, 
and the depopulation magnetic field is approximately $B_d \simeq 10 ~{\rm T}$.
Therefore, we obtain that the quadratic spin orbit and the Zeeman term cubic in $B$
have comparable strength: $|g_{03}|\simeq |g_{21}|\simeq 0.2$.

These estimates suggest that the apparent quenching of the many-body 
enhancement of the spin susceptibility is due to the presence of a 
large $n=2$ spin-orbit coupling. Moreover the presence of a sizable $n=3$ Rashba 
spin-orbit term would also result in a reduction of the generalized polarization.
This is consistent with the experimental finding that the suppression is most 
noticeable for $[100]$ quantum wells, for which the quadratic spin-orbit is 
comparatively stronger. 

From our theory one can moreover surmise that a larger many-body 
enhancement of the susceptibility is expected in the limit of very low 
densities when, due to the $\propto k^2$ and $\propto k^3$ dependence, 
the $n=2$ and $n=3$ terms become less relevant. 
The enhancement should be also more noticeable in the limit of a very 
narrow well. In the particular case of the $[113]$ growth direction, 
this happens because the linear Zeeman term (\ref{zeeman1}) can in principle become 
dominant. For the $[100]$ growth direction on the other hand, the term 
(\ref{zeeman1}) vanishes while the cubic Zeeman term (\ref{zeeman3}) is largest. 
This in spite of the $\sim W^4$ proportionality of the coupling strength. 
The reason is that the depopulation field is large in a very narrow well. 
This obtains since $B_d \sim 1/W^{\frac{4}{3}}$ while the magnitude of the 
quadratic spin-orbit (\ref{so2}) behaves like $\sim W^{\frac{2}{3}}$ thereby losing
its relevance.

It should be kept in mind that, being based on a perturbative treatment and 
not taking in account orbital effects, these conclusions should be taken at 
best as qualitative.\cite{commExchange5}

\section{Discussion and conclusions}
\label{disc}

The main conclusion of our analysis is that in the presence of quadratic and
cubic spin-orbit interactions the magnitude of the exchange energy decreases with
increasing generalized polarization. This results in a quenched value of $p$ and
in a corresponding increase of the value of the depopulation coupling $\bar\gamma_d$. 
By the same token the corresponding generalized susceptibility is also quenched.
This interesting phenomenon stems from the in plane rotation of the spin
quantization axis induced by the spin-orbit coupling proper ($n = 1,2,3$)
and from the universal structure of Eq.~(\ref{Ex}) which only depends on the
momentum space occupation (i.e. the generalized polarization $p$), and
the spin orientations $\hat s_{\bf k}$. As one can verify by making use of 
the very same equation, for a given value of $p$ the only difference between 
the various spin-orbit interactions stems from the different form acquired by 
$\hat s_{\bf k}$. Since for larger $n$ the spins are "less parallel" it is clear 
that the magnitude of the exchange energy will decrease for larger $n$.

Being based on a study of the exchange energy only, our theory can be expected to be 
strictly valid in the high density limit. The reason is that it is in this regime that the 
exchange energy represents the first interaction correction to the non interacting 
result, correlation effects becoming comparatively smaller as $r_s$ decreases.
On the other hand, at not too low densities, the physics of the exchange is still 
expected to give qualitatively reasonable results. This conclusion appears to be 
corroborated by the apparent observed reduction of the many-body enhancement of the spin 
susceptibility in dilute hole systems in which, beside the familiar Zeeman term, 
the magnetic field also induces large quadratic spin-orbit interactions. 

It is not difficult to prove that, when the spin-orbit coupling terms described 
by Eq.~(\ref{H0}) are present in isolation, the many-body states, as parameterized 
by $r_s$ and $p$, are also self-consistent solutions of the Hartree-Fock 
equations.\cite{gfg_proc05,unpub_ChesiGFG} 
On the other hand this is not the case when multiple concomitant terms are present. 
In such situations the circular symmetry is broken, and the interacting problem 
is considerably more complicated. Such situations must be treated case by case. 
In particular the spin quantization directions $\hat s_{\bf k}$ must be determined 
self-consistently.\cite{commExchange3} This problem has nontrivial solutions 
in the case of broken symmetry states, like for instance in the case of 
ferromagnetic phases.\cite{gfg_proc05,unpub_ChesiGFG} 

As a final remark we stress that since the generalized fractional polarization 
does not correspond directly to an actual magnetization, 
strictly speaking, one cannot draw direct conclusions about the 
enhancement of the spin-spin response from measurements of the depopulation field $B_d$. 
In general, the spin susceptibility is enhanced by the exchange, in a way similar to the 
usual case without spin-orbit.\cite{unpub_ChesiGFG} The bare spin-spin susceptibility involves
the response to a pure $n=0$ perturbation which, from an experimental point of view,
is not straightforward to realize for the case of hole systems of Ref.~\onlinecite{winkler05}. 
In fact, as we have argued, the external magnetic field induces also a change of the $n=2$ 
spin-orbit coupling. 

A detailed study of the linear spin-spin response and of the phase diagram, within the 
framework of the Hartree-Fock approximation and in the presence of spin-orbit coupling, 
will be the subject of future publications. Interestingly, the transition to a ferromagnetic 
state occurs at densities that are in general \emph{larger} than in the absence of 
spin-orbit interaction.\cite{gfg_proc05,unpub_ChesiGFG} 

\appendix*
\section{}

We obtain and discuss here spin dependent contributions to the effective hamiltonian
appropriate to the highest two-dimensional heavy hole subband in the presence of an 
in-plane magnetic field ${\bf B} = B_x \hat x + B_y \hat y$ for a typical III-V 
semiconductor quantum well. The specific numerical value of the parameters will
chosen to be appropriate to the case of $GaAs$. 

We begin by approximately describing the motion of the holes in the bulk 
through the standard Luttinger hamiltonian,\cite{luttinger56} 
which in spherical approximation takes the form: 
\begin{eqnarray}
\label{luttingerHspherical}
\hat H_{h}=-\frac{1}{2 m_0}\left[(\gamma_1+\frac{5}{2}\tilde\gamma)\, \hat {\bf p}^2 - 
2 \tilde\gamma( \hat {\bf J}\cdot \hat {\bf p})^2  \right] ~,
\end{eqnarray}  
where $\hat J_i$ are $4\times 4$ spin-3/2 matrices and, for the moment, cubic corrections 
have been neglected. For $GaAs$ $\gamma_1=6.85$ and $\tilde\gamma=(\gamma_2+\gamma_3)/2=2.5$. 
Within this context the effect of the magnetic field can be described by
introducing the Zeeman hamiltonian 
$\hat H_Z=-2 \kappa \, \mu_B \, {\bf B} \cdot \hat {\bf J}$, where for $GaAs$ $\kappa=1.2$,
a convenient choice of the vector potential being provided by 
${\bf A}=z B_y \hat{x} - z B_x \hat{y}$. 

As a specific model case, we consider here the confinement associated with an infinite 
rectangular well of width $W$. The corresponding effective hamiltonian for holes in the 
highest two-dimensional subband can be written as:
\begin{equation}\label{2Dsubband}
\hat H_0= \mathcal{E}_0(\hat {\bf p})+ \delta \hat H_{21}+ \delta \hat H_{03} ~,
\end{equation}
where $\mathcal{E}_0({\bf p})$ is the subband energy dispersion for ${\bf B}=0$,
and $\delta \hat H_{21}$, $\delta \hat H_{03}$ are spin dependent terms associated 
with the external magnetic field with sub indices indicating the value of the integer 
$n$ and the power of their dependence on the magnetic field.
Their explicit form can be found by making 
use of perturbation theory\cite{WinklerSpringer03} in ${\bf B}$ and in 
the wave vector ${\bf k}$. For the first term we obtain:
\begin{equation}\label{so2}
\delta \hat H_{21}= 
\frac{a \mu_B  W^2  }{\pi^2 \hbar^2}  \, 
\frac{B_+ p_+^2 \hat\sigma_- + B_- p_-^2 \hat\sigma_+}{2}  ~,
\end{equation}
where $B_\pm =B_x \pm i B_y$ and the explicit form of the numerical coefficient is
given by $a =\frac{1024 \tilde\gamma^2}{9\pi^2(3\gamma_1+10\tilde\gamma)}-\frac{3 \kappa}{2}$.
For the second term we find:
\begin{equation}\label{zeeman3}
\delta \hat H_{03}= b \mu_B^3
\left( \frac{ m_0 W^2}{\pi^2 \hbar^2} \right)^2  \, 
\frac{B_+^3 \hat\sigma_- + B_-^3 \hat\sigma_+}{2}  ~,
\end{equation}
where $b=\frac{\kappa (\pi^2-6)}{2}-\frac{27 \tilde\gamma^2}{8(2\gamma_1+5 \tilde\gamma)}$.
We should remark that the present results do differ from the ones one would 
infer from the corresponding formulas appearing in Ref.~\onlinecite{WinklerSpringer03}.
For $GaAs$ we have $a \simeq-0.2$ and $b \simeq 1.5$. 

With a suitable spin rotation, these contributions can be both transformed to the form 
of generic spin-orbit defined in our model hamiltonian (\ref{H0}). In particular, 
$\delta \hat H_{21}$ and $\delta \hat H_{03}$ are of the type $n=2$ and  $n=0$ 
respectively. Because of the isotropy implied by the spherical approximation, the value 
of the corresponding coupling strength $\gamma$ extracted by comparison to (\ref{H0}) 
only depends on the magnitude of the magnetic field $B$ and is immediately found to be:
\begin{equation}\label{gamma21}
\gamma_{21}=\frac{a \mu_B  W^2  B}{\pi^2 \hbar^2} ~,
\end{equation}
and
\begin{equation}\label{gamma03}
\gamma_{03}=b \mu_B^3
\left( \frac{W^2 m_0}{\pi^2 \hbar^2} \right)^2 B^3 ~,
\end{equation}for $\delta \hat H_{21}$ and $\delta \hat H_{03}$ respectively.

For a given growth direction, these results can be extended beyond the spherical 
approximation to include the appropriate cubic anisotropy as for instance done 
in Ref.~\onlinecite{WinklerSpringer03}.
Following this procedure one then obtains in (\ref{2Dsubband}) an additional anisotropic linear 
Zeeman term. By choosing coordinates along the principal axes, this can be generally expressed as:
\begin{equation} \label{zeeman1}
\delta \hat H_{01}=\frac{\mu_B}{2}(g_x B_x \hat \sigma_x + g_y B_y \hat \sigma_y)  ~.
\end{equation}
As it turns out, this term vanishes (even beyond the spherical approximation) 
for the high symmetry growth directions $[100]$ and $[111]$. However, in the case of the 
$[113]$ growth direction, (\ref{zeeman1}) is non vanishing and the principal axes $x$ and 
$y$ are along the $[1 \bar 1 0]$ and $[33\bar 2]$ respectively. 
A perturbative estimate of the suitable Land\'{e} g-factors for the case of an infinite 
rectangular well gives\cite{WinklerSpringer03} $g_x \simeq -0.17$ and $g_y\simeq 0.41$.
The coupling strength $\gamma$, as defined in (\ref{H0}), of this $n=0$ 
term depends on the direction of the magnetic field, and is given by:
\begin{equation}\label{gamma01}
\gamma_{01}^{i}=\frac{g_{i} \, \mu_B}{2}B ~,
\end{equation}
for the particular case of an external field of magnitude $B$ along 
one of the two principal axes ($i \equiv x $ or $y$).

Finally, we mention that including a transverse electric field one can develop 
a perturbation theory in $\mathcal{E}_z$ and ${\bf k}$ to obtain a term corresponding 
to the $n=3$ spin orbit coupling appearing in (\ref{H0}). In this case the explicit 
form of the coefficient reads:
\begin{equation}\label{gammacubic}
\gamma=\frac{512 \, e  \, \tilde\gamma^2 \, W^4}
{9 \hbar^3 \pi^6\, (\gamma_1 + 2 \, \tilde \gamma)
(3 \gamma_1 - 10 \, \tilde \gamma)} \, \mathcal{E}_z ~.
\end{equation}
This again differs from the corresponding expression surmised from
Ref.~\onlinecite{WinklerSpringer03}.




\begin{thebibliography}{18}
\bibitem[{com({\natexlab{a}})}]{comm_neutralizingbackground}
\bibinfo{note}{In this expression standard terms associated with the
  neutralizing background have been omitted for simplicity.\cite{TheBook}}.
\bibitem[{com({\natexlab{c}})}]{commExchange2}
\bibinfo{note}{The form of the spin-orbit coupling we consider is obviously not
  valid at large $k$.}
\bibitem[{com({\natexlab{d}})}]{commExchange4}
\bibinfo{note}{We will assume that for the $n=1$ Rashba case both chiral bands
  are occupied. When this is not the case (as it can in principle occur at very
  low densities or large spin-orbit coupling) one can tackle the problem by
  introducing the device of generalized chirality as done in
  Ref.~\onlinecite{gfg_proc05}.}
\bibitem[{com({\natexlab{e}})}]{commExchange5}
\bibinfo{note}{As pointed out in Ref.~\onlinecite{winkler05} a potentially
  relevant phenomenon we have neglected is that of the effective mass
  enhancement associated with the magnetic field.}
\end{thebibliography}
\end{document}